\documentclass[a4paper]{ESASPStyle}
\usepackage{epsfig}

\def\Ad{A_{\rm d}}

\def\diff{{\rm d}}

\def\muHz{\mu{\rm Hz}}
\def\Xc{X_{\rm c}}
\def\Xs{X_{\rm s}}
\def\Zs{Z_{\rm s}}
\def\btaud{{\bar{\tau}}_{\rm d}}

\def\note #1]{{\Large\bf #1]}}

\def\figsize{9.2truecm}
\def\figsizelarge{17.0truecm}
\def\figsdir{./}
\def\figspace{\vskip -0.4truecm}

\def\tabspace{\vskip -0.2truecm}

\begin{document}

\title{Asteroseismic Inference for Solar-Type Stars}

\author{M.J.P.F.G.~Monteiro\inst{1} \and J.~Christensen-Dalsgaard\inst{2}
  \and M.J.~Thompson\inst{3}}
\institute{
Departamento de Matem\'atica Aplicada da Faculdade de Ci\^encias, and \\
Centro de Astrof\'{\i}sica da Universidade do Porto, Rua das Estrelas,
P-4150-762 Porto, Portugal
\and
Teoretisk Astrofysik Center,  Danmarks Grundforskningsfond, and \\
    Institut for Fysik og Astronomi, Aarhus Universitet, 
    DK-8000 Aarhus C, Denmark
\and
Department of Physics, Imperial College, London, England }

\maketitle 

\begin{abstract}
The oscillation spectra of solar-type stars may in the not-too-distant
future be used to constrain certain properties of the stars.
The CD diagram of large versus small frequency separations
is one of the powerful tools available to infer the
properties -- including perhaps masses and ages -- of stars which 
display a detectable spectrum of oscillation. Also,
the border of a convective region in a solar-type star gives rise to 
a characteristic periodic signal in the star's low-degree p-mode frequencies.
Such a signature contains information about the location
and nature of the transition between
convective and non-convective regions in the star.

In this work we address some of the uncertainties associated with the
direct use of the CD diagram to evaluate the mass and age of the star
due to the unknown contributions that make the stars different from the
evolutionary models used to construct our reference grid.
We also explore the
possibility of combining an {\em amplitude versus period} diagram with
the CD diagram to evaluate the properties of convective borders 
within solar-type stars.

\end{abstract}

\section{Introduction}
  
Seismology of solar-type stars besides the Sun is expected 
in the not-too-distant future to provide information of great 
relevance for understanding stellar evolution, and here
we address the topic of using the frequencies of oscillation of
a few models of solar-type stars in order to constrain their structure.

The asteroseismic HR diagram (also called the ``CD diagram'')
is one of the powerful tools available to infer the
properties of stars which display a detectable spectrum of
oscillation. The CD diagram utilises the so-called
large and small frequency separations (defined in Section~2),
which are relatively easily identifiable characteristics of 
solar-type stars. Together they may provide information on a 
star's mass and age (e.g. Christensen-Dalsgaard 1993b).
However, as noted by Gough (1987), such inferences are affected by
other unknown aspects of the stellar properties; 
also, the errors in the observed frequencies must evidently be
taken into account.

Another aspect of a solar-type star's frequency spectrum is that the
border of a convective region (or indeed any other sharp variation)
gives rise to a characteristic periodic signal in the
low-degree p-mode frequencies.

In this work we use a grid of reference models both to construct
a CD diagram and
to calibrate the behaviour of the oscillatory signal from
the base of the convective envelope (the amplitude versus period diagram)
for stars of different masses and ages.
Further, we address some of the uncertainties associated with the
direct use of the CD diagram by considering grids of stellar models
with different input physics.
To illustrate the combined application of these tools, and to test the 
reliability of our inferences, in this paper we perform a blind 
test on frequency data from three stellar models, supplied by one
author to another.

\section{Properties of the frequencies}
\label{sec:separations}

The easiest information to measure from the frequencies of oscillation of
other stars is the {\em large frequency separation}.
This quantity is the regular spacing between modes of same degree
and consecutive order,
\begin{equation}
\Delta\nu_{n,l} \equiv \nu_{n{+}1,l}-\nu_{n,l}\;,
    \label{eq:large_sep}
\end{equation}
where $\nu_{n,l}$ is the frequency of the mode of order $n$ and degree $l$.
The large separation is mainly a measurement of the sound travel time between
the surface of the star and the centre:
\begin{equation}
2 \int_0^R {\diff r \over c}\;;
\end{equation}
however $\Delta\nu_{n,l}$ is a function of frequency.
Other aspects
also contribute to its value, besides the different sensitivity of
the modes to the global structure of the star.

 \begin{figure}[ht]
  \begin{center}
     \epsfig{file=\figsdir/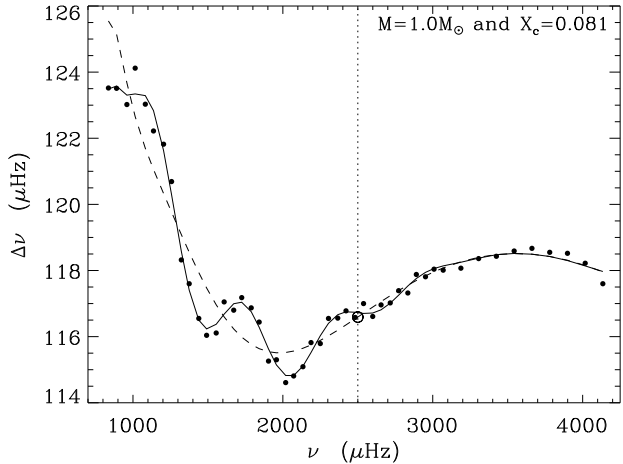, width=\figsize}
  \end{center}
  \figspace
  \caption{Large frequency separations for one model.
The continuous line is a fit to the points, while the dashed line
is the component of that fit that corresponds to powers of $(1/\nu)$. 
The difference is the contribution $\Delta\nu_b$
to the frequency differences due
to the presence of the second helium ionization zone near the
surface of the star.
The value at $\nu{=}2500\muHz$ has been defined
as the representative value of the separation and is indicated
as an open circle.
  \label{fig:large-sep}}
\end{figure}

 \begin{figure}[ht]
  \begin{center}
     \epsfig{file=\figsdir/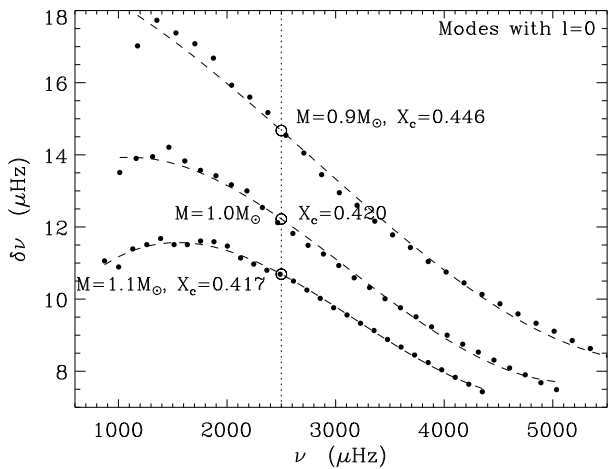, width=\figsize}
  \end{center}
  \figspace
  \caption{Small frequency separations $\delta\nu_{n,0}$
  for three models.
  The dashed line corresponds to a fit of the points with an expansion
  in powers of $1/\nu$. In order to find the value of $\delta\bar\nu$
  we use the value of the fit at $\nu{=}2500\muHz$.
  \label{fig:small-sep}}
\end{figure}

 \begin{figure}[ht]
  \begin{center}
     \epsfig{file=\figsdir/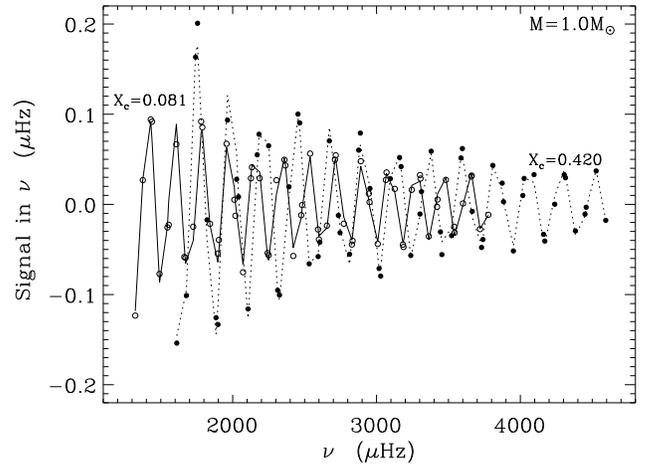, width=\figsize}
  \end{center}
  \figspace
  \caption{Signal in the frequencies due to the border of the
  convective envelope in two models of different ages for a star
  of a solar mass.
  \label{fig:signal} }
\end{figure}

The {\em small frequency separation} is defined according to
\begin{equation}
\delta\nu_{n,l} = \nu_{n,l} - \nu_{n{-}1,l{+}2} \;.
\end{equation}
The small separations are sensitive to evolutionary changes in the
structure of the core.
That makes this quantity sensitive to the age of the star as determined
by the central hydrogen abundance $\Xc$.

Sharp variations of the structure produce in the modes of oscillation a
characteristic perturbation that is proportional to the amplitude of
the mode at that location. 
This corresponds to a signal superimposed in the frequencies which
is periodic and with the amplitude measuring the sharpness of the transition.
Of particular note for solar-type stars
are the signal from the base of the convective envelope and that from
the second helium ionization zone.
Such a signal has been used with success for studying the base
of the convection zone in the Sun (\cite{basu94}, \cite{monteiro94},
\cite{roxburgh94}; \cite{jcd95}) 
and it can also be used for solar-type stars (\cite{monteiro00}).
The expression of the signal 
from one such transition layer
for very low degree modes can be
written as (see \cite{monteiro00})
\begin{equation}
    A(\omega)
    \cos\left[ 2(\omega\bar\tau_{\rm d} + \phi_0) \right]\;,
    \label{eq:signal}
\end{equation}
where $\omega$ is the frequency of the mode ($\omega{=}2\pi\nu$),
$A(\omega)$ the amplitude
of the signal, $\bar\tau_{\rm d}$ an acoustic depth and $\phi_0$ a phase.
The quantities $A(\omega)$ and $\bar\tau_{\rm d}$
are determined by the sharpness
of the transition and its acoustic location, respectively.

When determining the large separation, care should be taken to 
take account of 
additional contributions which are not included in the asymptotic
analysis.
Therefore we have removed the major contribution which
dominates at the lower range of frequencies (low mode order)
due to the presence of the second helium ionization
(e.g. \cite{monteiro98},
P\'erez Hern\'andez \& Christensen-Dalsgaard 1998).
We have done so by fitting to $\Delta \nu$ the sum of
a smooth component $\Delta \nu_a(\nu)$ and a component of the form
\begin{equation}
\Delta\nu_{b} \propto  {\sin^2(\beta\omega) \over \omega} \;
  \cos\left[ 2(\bar\tau_d \omega + \phi_0)\right]  \; ,
\end{equation}
which is a special case of an oscillatory signal as described by
Eq.~(\ref{eq:signal}).
Here the quantities $\beta$ and $\bar\tau_d$
characterise the second helium ionization zone
(width and acoustic depth respectively - see \cite{monteiro98}).

Figure~1 illustrates our method:
the smooth component $(\Delta\nu_a)$ is shown as a dashed
line, and results from a fit of the form
$\Delta\nu{\equiv}\Delta\nu_a{+}\Delta\nu_b$.
In order to calibrate the dependence of the large separation
on the global properties of the star
we consider its value at $\nu{=}2500\muHz$.
We keep only the smooth component to calculate
\begin{equation}
\Delta\bar\nu \equiv \Delta\nu_a(\nu{=}2500\muHz) \; .
\end{equation}

Figure~2 shows the small frequency separations for $l{=}0$
and the fit we consider in order to define the value at 2500$\muHz$ as:
\begin{equation}
\delta\bar\nu \equiv \delta\nu(\nu{=}2500\muHz) \; .
\end{equation}

 \begin{figure*}[ht]
  \begin{center}
     \epsfig{file=\figsdir/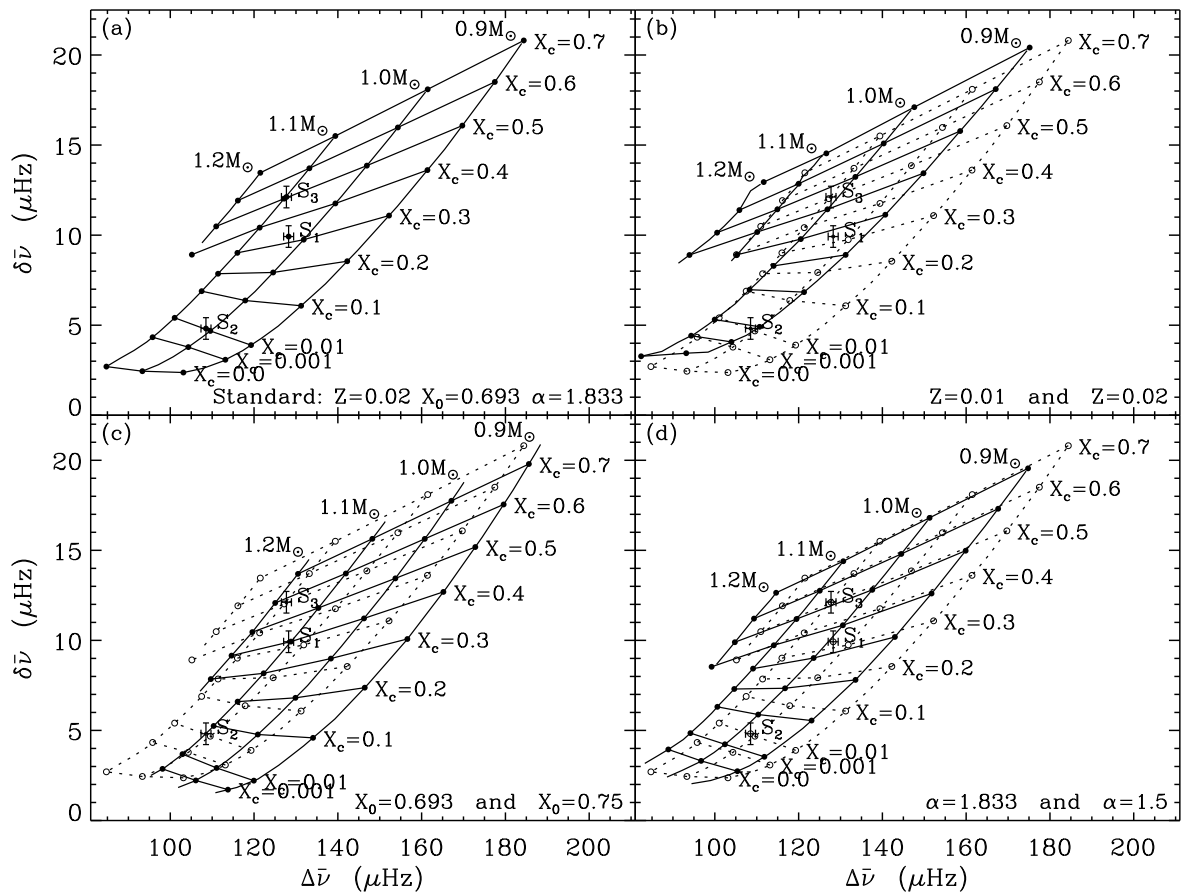, width=\figsizelarge}
  \end{center}
  \figspace
  \caption{(a) The CD diagram for evolutionary sequences of
  0.9, 1.0, 1.1 and 1.2 solar mass stars - all the models have $Z{=}0.02$,
  $X_0{=}0.693$ and $\alpha {=} 1.833$ (see Section~\ref{sec:cd-diagram}).
  The test stars (see Table~\ref{tab:data})
  are also indicated in all panels with 3-sigma error bars.
  (b) CD diagram for models with $Z{=}0.01$ (continuous line).
  The ``standard'' diagram from panel (a), with
   $Z{=}0.02$ (dotted lines) is shown as a reference.
  (c) CD diagram for models with $X_0{=}0.75$ (continuous line).
  The ``standard'' diagram with $X_0{=}0.693$ (dotted lines) is also shown.
  (d) CD diagram for models with $\alpha {=} 1.5$ (continuous line).
  Our ``standard'' diagram with $\alpha {=} 1.833$ (dotted lines) is also
  shown.
  \label{fig:cd-diagram}}
\end{figure*}

Figure~\ref{fig:signal} illustrates the oscillatory signal from the 
base of the convection zone in two stellar models. The signal is
extracted from the frequencies in the manner described by \cite*{monteiro94} 
and \cite*{monteiro00}.
For comparing the properties of convective borders of stars of
different age and mass we 
use the value of the signal amplitude at frequency
$\tilde\omega/2\pi{=}2500\muHz$, by defining $\Ad{\equiv}A(\tilde\omega)$.
The measured quantities $\Ad$ and $\btaud$ 
provide direct constraints on the 
stratification and location of the base of the convective envelope of the star.

\section{Dependence on stellar mass and age}

Evolutionary sequences of different-mass stars
have been calculated as described by \cite*{jcd93a}.
In particular, the EFF equation of state (\cite{eggleton73})
was used, as well as OPAL92 opacities (\cite{rogers92}).

\subsection{The CD diagram}
\label{sec:cd-diagram}

{}From a set of evolutionary sequences of stellar models
we compute the stars' eigenfrequencies and hence position each star on
a ($\Delta\bar\nu,\delta\bar\nu)$ diagram - the so-called CD Diagram.
As a reference case, we use a ``standard'' set of parameters
($Z{=}0.02$, initial hydrogen abundance $X{=}0.693$, mixing-length parameter
$\alpha {=} 0.183$); the resulting diagram is shown in
Fig.~\ref{fig:cd-diagram}($a$). 
A star's position in the diagram depends on its mass and its age.

Any unknown star whose large and small separations have been calculated
in the same way can now be located within this diagram and hence its
mass and age can be estimated. This can be done independently of any
other observational information about the star.
But any property which is known for
the models in the calibration set may be estimated similarly, e.g. luminosity 
or effective temperature. Of course all such calibration may
be inaccurate if the other physics of the star is unlike that used in
the models used to construct the diagram. Therefore we consider the 
effect on the CD diagram of modifying aspects of the physics of the
stellar models, specifically the metal abundance $Z$, the hydrogen
abundance $X$, and the mixing length parameter $\alpha$.

Figure~\ref{fig:cd-diagram}($b$) shows the CD diagram obtained using 
models with a heavy element abundance $Z {=} 0.01$, 
the other parameters having the ``standard values''.
Figure~\ref{fig:cd-diagram}($c$) shows the CD diagram obtained using
models with an initial hydrogen abundance
$X_0 {=} 0.75$, the other parameters have their standard values.
Finally, Fig.~\ref{fig:cd-diagram}($d$) illustrates the effect of 
using a mixing-length parameter $\alpha{=} 1.5$ with other parameter
values as standard.

Calibrating with each such CD diagram will in general give a different
estimate of the mass and age of the star.  However, if other
quantities are also known independently of seismology for the unknown
star, such calibration provides a consistency check and may reveal
inconsistencies between the unknown star and the physics assumed for
the calibration models.

\subsection{Convective borders}
\label{subsec:mass-age}

\begin{figure}[ht]
  \begin{center}
    \leavevmode
  \centerline{\epsfig{file=\figsdir/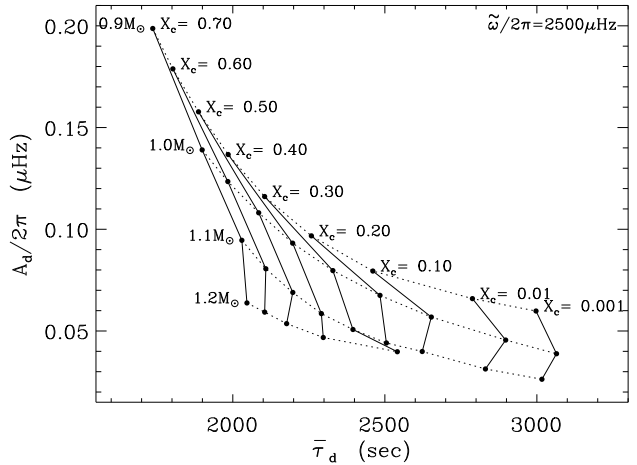,width=\figsize}}
  \end{center}
  \figspace
  \caption{Diagram of the amplitude of the signal versus
  the acoustic depth locating the base of the convection zone for
  our `reference' set of models.
  The dotted lines indicate models of the same mass while continuous
  lines are models with the same central hydrogen abundance
  $X_{\rm c}$, as indicated.}
  \label{fig:amp-tau}
\end{figure}

The amplitude and acoustic-depth parameters deduced for 
a convective border also depend on the mass and age of the star.
In order to illustrate the dependence on mass and age we
consider again the same sequence of models of different masses evolved
from the ZAMS to the end of the main sequence
(see above).
Here we assume that the
central helium abundance $X_{\rm c}$ is a measure of the age of
the star in the main sequence.
No type of overshoot has been included in this set of
evolutionary sequences.
In Fig.~\ref{fig:amp-tau} the behaviours of both parameters are
shown for all models.

\begin{figure}[ht]
  \begin{center}
    \leavevmode
  \centerline{\epsfig{file=\figsdir/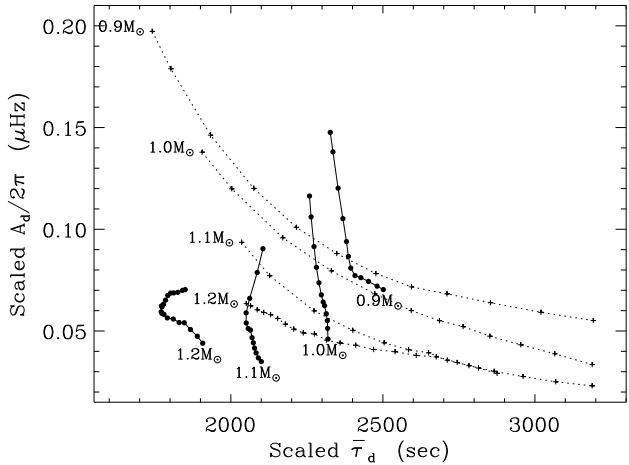,width=\figsize}}
  \end{center}
  \figspace
  \caption{Diagram of $\Ad{\times}\sqrt{R^3/M}$
  versus $\btaud{\times}\sqrt{M/R^3}$,
  where $R$ and $M$ are surface radius and total mass of the stars,
  in solar units.
  The dotted lines show the original values, without 
  the homology scaling factor.
  The plot illustrates how much of the variation of these two parameters
  is not homologous.
  Note that the amplitude decreases with increasing age.}
  \label{fig:amp-tau-scale}
\end{figure}

A large part of the variation of the large separation, the amplitude
and the acoustic depth arise from the variation of the mass $M$ and
radius $R$ of the stars whereby quantities with dimensions of frequency 
vary as $\sqrt{M/R^3}$. (The variation of the small separation is
affected more by nonhomologous variations in structure.)
To illustrate that, in Fig.~\ref{fig:amp-tau-scale}
we rescale the amplitude and acoustic depth 
\begin{equation}
\Ad \times \left[ {M/M_\odot\over\left( {R / R_\odot}\right)^3}
      \right]^{-1/2}
\qquad {\rm and}\qquad
\btaud \times \left[ {M/M_\odot\over\left( {R / R_\odot}\right)^3}
      \right]^{1/2}
    \label{eq:scales}
\end{equation}
to reveal other variations due to the changing
size of the envelope, which affects not only the
location but also the sharpness of the boundary.

\begin{table*}[htb]
  \begin{center}
    \caption{Data provided for the seismic analysis of the stars.
    The values of the mode degree $l$ and order $n$ for all
    frequencies considered are listed,
    to show what frequency data we have considered. 
    In order to simulate a real observation we also consider as known
    the luminosity $L$ of the star, its effective
    temperature $T_{\rm eff}$ and its metallicity $\Zs/\Xs$.}
    \leavevmode \tabspace
    \begin{tabular}[h]{lccccccccc}
      \hline \\[-5pt]
      Star & $l$ & $n$ & $\sigma(\nu)$ ($\mu$Hz)&
             $L/L_\odot$ & $T_{\rm eff} (K)$ & $\Zs/\Xs$  \\[+5pt]
      \hline \\[-5pt]
      $S_1$  & 0 & 14-32 & 0.1 & 1.186$\pm$0.036 & 5810.6$\pm$50.0 
          & 0.0274$\pm$0.0014 \\
          & 1 & 13-29 \\
          & 2 & 16-28 \\[+5pt]
      $S_2$  & 0 & 14-32 & 0.1 & 1.355$\pm$0.041 & 5725.0$\pm$50.0 
          & 0.0229$\pm$0.0011 \\
          & 1 & 13-29 \\
          & 2 & 15-30 \\
          & 3 & 16-28 \\[+5pt]
      $S_3$  & 0 & 14-32 & 0.1 & 1.259$\pm$0.038 & 5947.9$\pm$50.0 
          & 0.0282$\pm$0.0014 \\
          & 1 & 13-29 \\
          & 2 & 15-30 \\
          & 3 & 16-28 \\[+5pt]
      \hline
      \end{tabular}
    \label{tab:data}
  \end{center}
  \tabspace
\end{table*}

\begin{table*}[htb]
  \begin{center}
    \caption{Mass and central hydrogen abundance for the test
    stars as found from the CD diagrams shown in Fig.~\ref{fig:cd-diagram}
    (a), (b), (c) and (d) respectively. Also given are the values of
    the luminosity and effective temperature inferred from the 
    same sets of calibration models.}
    \leavevmode  \tabspace
    \begin{tabular}[h]{lcclccccccc}
      \hline \\[-5pt]
      Star & $\Delta\bar\nu$ $(\muHz)$ & $\delta\bar\nu$ $(\muHz)$
           & Physics &  $M/M_\odot$ & $\Xc$
           &  $L/L_\odot$(cal) & $T_{\rm eff}$(cal) \\[+5pt]
      \hline \\[-5pt]
      $S_1$ & 128.3 &  9.92 & $Standard$
          & 1.035$\pm$0.009 & 0.330$\pm$0.018 
          & 1.166 & 5859 \\
            &       &       & $Z{=}0.01$ 
          & 0.949$\pm$0.009 & 0.283$\pm$0.017 
          & 1.279 & 6085 \\
            &       &       & $Y{=}0.75$ 
          & 1.104$\pm$0.010 & 0.403$\pm$0.019 
          & 1.108 & 5729 \\
            &       &       & $\alpha{=}1.5$ 
          & 0.992$\pm$0.009 & 0.347$\pm$0.015 
          & 0.948 & 5606 \\[+5pt]
      $S_2$ & 108.5 &  4.82 & $Standard$
          & 1.014$\pm$0.012 & 0.016$\pm$0.007 
          & 1.444 & 5836 \\
            &       &       & $Z{=}0.01$ 
          & 0.910$\pm$0.012 & 0.011$\pm$0.007 
          & 1.517 & 6014 \\
            &       &       & $Y{=}0.75$ 
          & 1.099$\pm$0.012 & 0.067$\pm$0.014 
          & 1.436 & 5759 \\
            &       &       & $\alpha{=}1.5$ 
          & 0.975$\pm$0.011 & 0.045$\pm$0.009 
          & 1.202 & 5600 \\[+5pt]
      $S_3$ & 127.7 & 12.12 & $Standard$
          & 1.098$\pm$0.010 & 0.504$\pm$0.023 
          & 1.328 & 5996 \\
            &       &       & $Z{=}0.01$ 
          & 1.015$\pm$0.010 & 0.457$\pm$0.023 
          & 1.460 & 6231 \\
            &       &       & $Y{=}0.75$ 
          & 1.177$\pm$0.010 & 0.585$\pm$0.020 
          & 1.286 & 5865 \\
            &       &       & $\alpha{=}1.5$ 
          & 1.060$\pm$0.010 & 0.526$\pm$0.020 
          & 1.104 & 5746 \\[+5pt]
      \hline
      \end{tabular}
    \label{tab:res-all}
  \end{center}
  \tabspace
\end{table*}

\section{Test stars - observational data}
\label{sec:observations}

By way of illustration we consider three test stars which were provided 
from one co-author to another as a blind test.
Table~\ref{tab:data} lists all the information and errors bars that were
assumed as known for the test stars.

\subsection{The CD-Diagram fits}

We can use the procedure described in Section 3 
to analyse these data in terms of our CD diagrams;
the positions of the stars have been shown in 
Fig.~\ref{fig:cd-diagram}.
The inferred values of the stellar parameters
are given in Table~\ref{tab:res-all}.
The errors bars have been estimated by determining the uncertainties
due only to the observational uncertainties for the frequencies.
As well as $M$ and $\Xc$, the effective temperature and luminosity
of the star can be calibrated from the evolutionary tracks corresponding
to the CD diagram.  These values are also shown in Table~\ref{tab:res-all}.

\subsection{Fits for the convective borders}

By isolating the signal in the frequencies, with a fit of the
expression (\ref{eq:signal}) we determine a value for the
amplitude $\Ad$ and the acoustic location $\btaud$.
The results are given in Table~\ref{tab:signal}.
The error bars have been estimated in accordance with \cite*{monteiro00}
from the error of $0.1\muHz$ in the individual frequencies.


\begin{table}[htb]
  \begin{center}
    \caption{Parameters of the signal as found from fitting
    the expression of the signal to the frequencies of the stars.}
    \leavevmode \tabspace
    \begin{tabular}[h]{lccccccc}
      \hline \\[-5pt]
      Star & $\btaud$ (sec)& $\Ad/2\pi$ ($\mu$Hz) \\[+5pt]
      \hline \\[-5pt]
      $S_1$  &  2201$\pm$45  &  0.103$\pm$0.014 \\
      $S_2$  &  2404$\pm$42  &  0.045$\pm$0.008 \\
      $S_3$  &  2272$\pm$73  &  0.164$\pm$0.012 \\[+5pt]
      \hline
      \end{tabular}
    \label{tab:signal}
  \end{center}
\tabspace
\end{table}

\section{Combined inferences}

With the values found above we can now attempt to estimate what
type of stars we have as blind tests.

\subsection{Masses and ages}

The first point to note is that the masses and central hydrogen
abundances inferred for our test stars, given in Table~\ref{tab:res-all},
are discrepant. This shows the sensitivity of the calibration based on
the frequency separations to the physics assumed when
the CD diagram is constructed. From the large and small separations 
alone there is no way to discriminate between these different inferences. 

\begin{figure}[ht]
  \begin{center}
      \epsfig{file=\figsdir/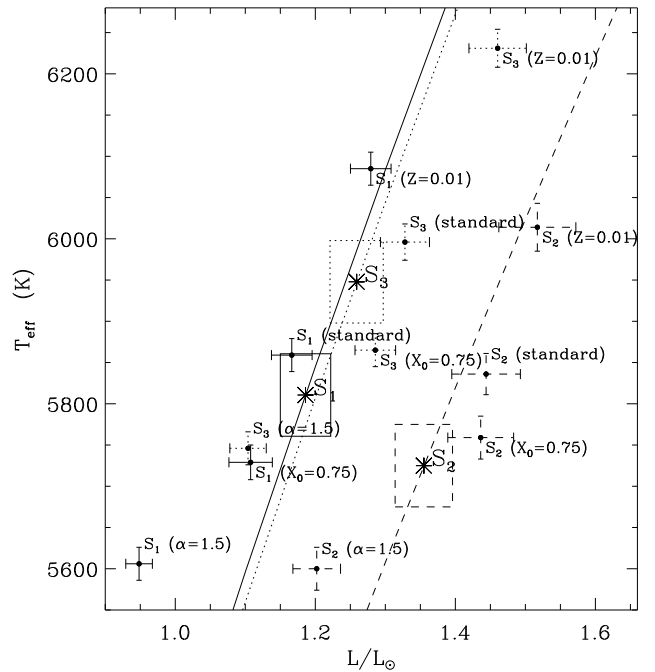, width=\figsize}
  \end{center}
  \figspace
  \caption{ Luminosity (in solar units) and effective temperature
  for the test stars.
  The large symbols with error boxes show the `observed' values
  provided in Table~\ref{tab:data}.
  The remaining points have been determined from the inferred
  $(M, \Xc)$ for each of the four sets of reference models,
  using the appropriate evolutionary tracks.
  The standard calibration case has $Z{=}0.02$, $X_0{=}0.693$
  and $\alpha{=}0.183$.
  The lines indicate models of constant radius, as
  determined from the observations of the luminosity and
  effective temperature for each test star.
  \label{fig:teff-l}}
\end{figure}

However, for our test stars, and we expect for future asteroseismic
target stars
also, we have additional observational constraints. In the present case
we have independent estimates of the luminosities and temperatures of the 
stars. These quantities can also be estimated for the calibration sets and
hence internal inconsistencies may be revealed. 
Specifically, we show in Fig.~\ref{fig:teff-l}
the specified locations of the test stars (cf. Table~\ref{tab:data})
with the associated error boxes;
in addition, the figure shows those values of $T_{\rm eff}$ and $L$
that would be obtained from the values of $M$ and $\Xc$ inferred
from the four CD diagrams in Fig.~\ref{fig:cd-diagram}
(see Table~\ref{tab:res-all}).
This allows us to test the consistency of our inferences.

As given in Table~\ref{tab:data} we also have the values
of $\Zs/\Xs$ for each star.
This quantity complements Fig.~\ref{fig:teff-l} by providing
an extra consistency test.

Using our standard CD diagram (Fig.~\ref{fig:cd-diagram})
gives consistent values for
$S_1$ (compare values of $L$ and $T_{\rm eff}$ in
Tables~\ref{tab:data} and \ref{tab:res-all}).
We conclude that
star $S_1$ has a mass $M/M_\odot{=}1.035{\pm}0.009$, and a central hydrogen
abundance of $\Xc{=}0.330{\pm}0.018$.
The star appears to be consistent with our calibration models, which have
a heavy element abundance of $0.02$
and an initial abundance of hydrogen of $0.6928$.
We note that the specified value of $\Zs/\Xs$ is consistent with this
conclusion.\\

For case $S_2$,  the luminosity obtained by calibration to our standard CD
diagram (Table~\ref{tab:res-all}) is larger than the star's
observed luminosity (Table~\ref{tab:data}).
Likewise the effective temperature is discrepant.
We conclude that the physics and/or abundances of $S_2$
are not the same as those used in the calibration models.
If instead we use $Z{=}0.01$, the calibrated 
values of $L$ and $T_{\rm eff}$ are even more discrepant.
In order to reconcile the values we can either
increase the value of $Z$ or increase the value of $X_0$.
Taking into consideration the measured metalicity $\Zs/\Xs$
the best option would be to increase $X_0$.
We can also decrease slightly the value of $\alpha$,
corresponding to saying that this star has a higher value
of $X_0$ and a lower value of $\alpha$.
We clearly need to increase the radius (and hence, by homology,
also increase the mass) of the star
that comes out of the standard CD diagram
(Fig.~\ref{fig:cd-diagram}).
Taking these factors in consideration we would say the global properties
for $S_2$ are that $M/M_\odot{=}1.05{\pm}0.01$ and $\Xc{<}0.1$.
For this star we should build first a
more adequate CD diagram before calculating its global properties.

For case $S_3$ the value of the luminosity obtained by our standard
calibration (Table~2) is again higher than actually observed, though
the calibrated and observed temperatures are marginally consistent within the 
observational error bars.
This star seems to indicate that it has a high $\Zs/\Xs$.
It probably also has a higher value of $X_0$
which is associated with a lower value of $\alpha$.
That is, we would say this star has $Z{=}0.021$ and $X_0{\simeq}0.72$.
If that is the case then we may propose that this star has
$M/M_\odot{=}1.1{\pm}0.02$ and $\Xc{\simeq}0.55{\pm}0.02$.
Again, we should iterate on the physics in order to
get a better CD diagram for this star.


\subsection{Convective borders}

\begin{figure}[ht]
  \begin{center}
    \leavevmode
  \centerline{\epsfig{file=\figsdir/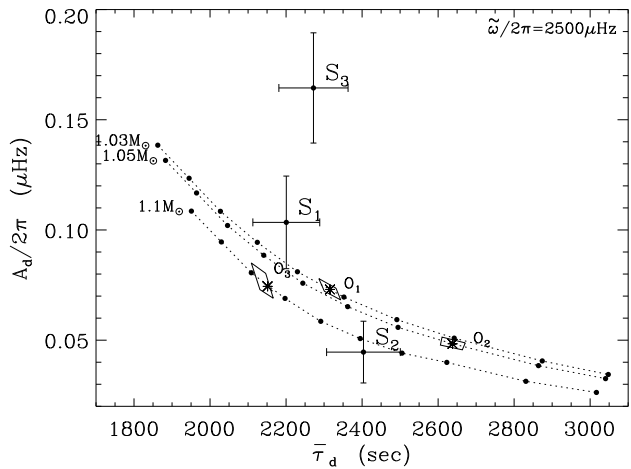,width=\figsize}}
  \end{center}
  \figspace
  \caption{Diagram of the amplitude of the signal versus
  the acoustic depth locating the base of the convection zone for
  our test stars ($S_1$, $S_2$ and $S_3$). The values we would
  expect to measure in models with no overshoot,
  considering the mass and age we have previously
  determined for these stars, are also indicated
  ($O_1$, $O_2$ and $O_3$).
  }
  \label{fig:amp-tau-stars}
\end{figure}

Assuming that we have been successful with the determination of the
stellar mass and central helium abundance, we now
try to determined the main characteristics of these stars at the base of 
their convective envelopes. 
Fig.~\ref{fig:amp-tau-stars} shows the measured 
quantities $\Ad$ and $\btaud$ 
for the test stars, and the values expected for our reference set
of models.
For comparison we also show what values we would expect 
in the reference models with no overshoot, for stars with masses and
ages as determined for $S_1$, $S_2$ and $S_3$ 
from the CD diagram analysis. Note that, rather than 
using our reference-case models, it would be more consistent to
calibrate against 
model sequences computed according to our best-fit parameters determined
in Section~5.1.

Both models $S_1$ and $S_3$ show a higher amplitude than expected and thus
appear to differ from the corresponding references models in terms of
the stratification at the base of the convective envelope.
An adiabatically stratified overshoot layer
of $0.2H_p$ for $S_3$ and $0.1H_p$ for the less massive star $S_1$ could
account for this.
These limits were calculated using the dependence of the
amplitude on overshoot, for different ages, as given by \cite*{monteiro00}.
For $S_1$ the measured value of $\btaud$ is smaller than expected,
which is not what we should see with the presence of an overshoot layer.
That may indicate that the sharpness we measure in the amplitude
could be associated with diffusion at the base of the envelope,
and not overshoot as said above. (In the blind test, the investigator
did not provide error bars on the above-quoted overshoot values. In fact
we note a posteriori that the remarked-upon deviation of $S_1$ from
the corresponding reference model is only at the 1-$\sigma$ level.)

Regarding $S_2$ there is no difference in amplitude indicating that
the smoothness of the transition 
in $S_2$ is similar to that in the reference models.
However, there is an indication of a smaller acoustic depth (at the 
2-$\sigma$ level), which we
could attribute to the atmospheric contribution to $\btaud$
(a different reflecting layer at the top of the star will
introduce a shift in $\btaud$ of up to about 100 sec).
Overshoot that is not adiabatically stratified can
also introduce a signal which differs from the expected values
mainly in the acoustic depth, and not in amplitude.
But such a fuzzy extension by overshooting 
would tend to increase $\btaud$ relative
to our reference models, and so does not appear to be relevant here.

\section{The truth}
\label{sec:truth}

After the completion by one of the authors of the preceding analysis
of the test data, the true properties of the underlying
models, presented in Table~\ref{tab:truth}, were revealed.
The following additional comments about the models may be made:

{\bf $S_1$} -
This was constructed using the same physics and parameters
as our `standard' reference set.
As concluded, the inference of $M$ and $\Xc$ was quite secure.\\

{\bf $S_2$} -
This model used the OPAL equation of state
(Rogers, Swenson \& Iglesias 1996) and OPAL95 opacities
(Iglesias \& Rogers 1996), and included diffusion and
settling of helium and heavy elements.\\

{\bf $S_3$} -
This used the same physics and parameters
as our `standard' reference set, except for including overshoot
from the convective envelope, over 0.2 pressure scale heights.

\begin{table*}[htb]
  \begin{center}
    \caption{True masses and central hydrogen abundances for the test
    stars for which inferences were attempted.
    Also indicated are
    the size of the overshoot layer ($\ell_{\rm ov}$)
    in units of the local pressure scale-height ($H_p$),
    which has been modelled as being adiabatic,
    and whether or not the star has settling/diffusion (Dif.) and uses
    different opacity tables and equation of state (Phys.)
    from our reference models.}
    \leavevmode   \tabspace
    \begin{tabular}[h]{lccccccc}
      \hline \\[-5pt]
      Star & $\Delta\bar\nu$ $(\muHz)$ & $\delta\bar\nu$ $(\muHz)$ &
             $M/M_\odot$ & $\Xc$ & $\ell_{\rm ov}/H_p$ & Dif. &
             Phys. \\[+5pt]
      \hline \\[-5pt]
      $S_1$  & 128.3 &  9.92 & 1.033 & 0.333 & 0.0 & no  & no \\
      $S_2$  & 108.5 &  4.82 & 1.000 & 0.017 & 0.0 & yes & yes\\
      $S_3$  & 127.7 & 12.12 & 1.100 & 0.521 & 0.2 & no  & no \\[+5pt]
      \hline
      \end{tabular}
    \label{tab:truth}
  \end{center}
  \tabspace
\end{table*}

\section{Discussion}
\label{sec:discussion}

Our process of inference proceeds in two steps: firstly, estimating
the stars' masses and ages with the CD diagram, and then investigating
the sharpness and location of the convective borders.

\subsection{The CD-Diagram}

We have demonstrated that the physics adopted for the models 
used to construct the CD diagram has an effect on the mass and
central hydrogen abundance inferred from the large and small separations.
We have quantified this effect for three possible changes: the 
heavy-element abundance, the initial hydrogen abundance; and the 
mixing-length parameter. We have also quantified the uncertainties
in the seismic determinations due to observational errors in the
frequencies.

We have shown how additional observational data
(specifically a star's luminosity and effective temperature) 
complement the seismic data and thus enable the mass and 
central hydrogen abundance to be better determined. Such additional
constraints may also permit inconsistencies in the physics to be 
revealed.
Thus
introducing information from additional observables (luminosity, 
effective temperature, chemical abundances, surface gravity, ...) can
raise the degeneracy that otherwise exists when attempting 
to calibrate stars using just two seismic measurements, the large
and small separations (e.g. \cite{brown94}).

Of course it should also be borne in mind
that the frequency separations actually depend on frequency (e.g. Figs~1 
and 2) and are not simply two numbers: this variation also contains 
valuable information about the objects of study which we have not
as yet exploited in this study.

\subsection{The convective borders}

We have had fair success in the first step.
That is very important in order to be able to perform the comparison of
the amplitude  and the acoustic depth of the signal with a set of
models of solar-type stars of different ages, as we have done here.

We anticipate that difficulties of interpretation in the second step
may arise from the opposing contributions of
different aspects of the physics relevant for the base of the envelope.
The frequencies only provide us a measure of the sharpness of that transition.
This aspect is dependent on every aspect of the physics at play at
that location, including convection and overshoot, the equation of state,
opacities, diffusion and settling,  as well as aspects not included in
our present models but undoubtedly at play in real stars 
(magnetic fields, rotation, etc.).
Therefore to translate some of the values we find here for the
amplitude we must make sure that we have reproduced as closely as
possible in our models the physics of the star.
That also includes the CD diagram analysis where such an
iteration must be done.
For a star with a large adiabatically stratified overshoot layer we
had no problems identifying its presence ($S_3$).
This information could and should be fed back into the CD Diagram
analysis in order to improve the estimates provided for the mass
and central helium abundance.

The inferences for the star $S_2$ with modified physics were 
also reasonably successful.
The initial guess was that the apparently slightly anomalous value of
$\btaud$ was due to the surface contribution.
In fact, settling, a different equation of state and different
opacity tables in this model 
have changed the expected value of $\btaud$, but we do have difficulties in
determining that without going back to the CD diagram where an
iteration on the physics could lead us to adapt the physics in order
to satisfy all observational constraints.

Finally, the case of $S_1$ was on the face of it a little disappointing.
The step of estimating the mass and age was very successful, and indeed
the star had input physics matching that of the grid of reference models.
But the inferences about the base of its convective envelope were
wide of the mark. 
The lesson, which is brought home most forcefully in a blind test,
is that one can overinterpret noisy data.
As we saw with hindsight, the inferred values of 
the amplitude and acoustic depth for this star are only just over one
standard deviation from the values expected from the reference
models.

\begin{acknowledgements}

This work was supported in part
by the Portuguese Funda\c{c}\~ao para a Ci\^encia e a Tecnologia,
by the Danish National Research Foundation through
   its establishment of the Theoretical Astrophysics Center
and by the UK Particle Physics and Astronomy Research Council.

\end{acknowledgements}


\end{document}